\documentstyle[pre,aps,psfig]{revtex}
\begin{document}
\title{Determination of the electrical conductivity and the electroosmotic
transfer in the concentrated dispersions on the basis of the cell theory 
of the electroosmosis}
\author{D.L.Gulo and O.L.Alexejev}
\address{F.D.Ovcharenko	Institute of Biocolloidal Chemistry , National Academy
of Sciences of Ukraine,\\42 Vernadsky blvd ., 252142 Kyiv , Ukraine}
\draft\date{\today}
\maketitle\begin{abstract}
Theoretical calculations of the electrical conductivity
and electroosmotic transfer as functions of the disperse phase volume fraction
and non-dissolving boundary layer thickness were provided on the basis of the 
cell theory of electroosmosis for the limiting case of large degree of electric
double layers' overlapping in interparticle space . The obtained results are
in qualitative agreement with the experimental data  and describe the main 
features of the latter .
\end{abstract}


\section{Introduction} 

The obtaining of equations , which can accurately describe electrokinetic
phenomena in concentrated dispersions , is impeded by a series of causes . 
The influence of the interaction between equilibrium and polarization fields
of colloid particles becomes considerable . The properties of disperse medium 
are also changed by the surface of solid phase . Liquid layers  being near 
the surface and having peculiar properties  occupy a substantial volume part 
of a concentrated dispersion threshold space and strongly influence the 
electrokinetic phenomena .Systematic investigations of the electrokinetic
phenomena in the systems of colloidal dispersity led to the discovery of
regularities [ 1- 3 ] which initiated the elaboration of  the electroosmosis
theory in plane capillaries for the limiting case of large degree of electric
double layers' overlapping [ 4 ] . Unlike the previous work [ 5 ] , the model 
of the near wall layer of fluid in which it combined the non - dissolving volume 
properties with the hydrodynamic mobility was developed and this gave a theoretical 
explanation for many experimental results . The continuation of investigations 
in this field was the construction of the cell theory of the electroosmosis [ 6 ] 
where the cell model widely applied while solving hydrodynamic tasks [ 7 ] 
was used to  electrokinetic problems .	The present paper contains theoretical
calculations of the conductivity and electroosmotic transfer as functions of
disperse phase volume fraction for experimentally investigated systems and 
the comparison of obtained calculations with the experimental data . 

\section{Statement and solution of a task} 
	In the cell model a single particle ( radius $a$ ) is considered to be 
surrounded by a sphere of such a size ( radius $c$ ) that the fraction of 
liquid within the sphere is the same as the porosity of the colloid system . 
Suspension is considered as the aggregate of these typical structural units. 
Having designated the volume particle fraction in the suspension by $\phi$ , we
obtain :
\begin{equation}
\phi = a^{3} / c^{3}.\eqnum{1}
\end{equation}                                                                                                
Having designated the radius of the dissolving and non - dissolving layer in 
the cell by $b$  , the thickness of the non - dissolving boundary layer $h$  is :
\begin{equation}
h = b - a, \eqnum{2}
\end{equation}
(see Fig.\ref{fig1}).		 	                                                           
In the work [ 6 ] for the calculation of the electroosmosis rate of the 1 - 1 
binary electrolyte the following equation system was solved :
\begin{equation}	 	
-\eta\;rot\,\/rot\;\vec{\nu}-grad\,(p+RT\gamma^{+}\mu^{+}+RT\gamma^{-}\mu^{-})=0, \eqnum{3}
\end{equation}
\begin{equation}
div\vec{\nu}=0, \eqnum{4}
\end{equation}
\begin{equation}
div\,grad\,\mu^{\pm}=0, \eqnum{5}
\end{equation}	 	
where $\eta$ is  the fluid viscosity, $p$ is the pressure, $\nu$ is the 
fluid rate, $\mu$ are dimensionless (in the $RT$ units) electrochemical
ion potentials, $C$ are their concentrations. In this case 
$\gamma{^\pm} = C - C_o$, $C_o$ is  the concentration of the equilibrium
solution of the electrolyte.
The general solutions were given by the equations :
\begin{equation}	 	
\mu^{\pm} =\cos\theta\,(A_{1}^{\pm} r+A_{2}^{\pm} r^{-2} ),\eqnum{6}
\end{equation}
\begin{equation}
\nu_{r} =\cos\theta\,(B_{1} -\frac{2B_{2} }{r^{3} } +\frac{B_{3} r^{2} }{10} +
\frac{B_{4} }{r} ), \eqnum{7}
\end{equation}
\begin{equation}
\nu _{\theta } =-\sin \theta \,(B_{1} +\frac{B_{2} }{r^{3} } +\frac{B_{3}r^{2} }{5} +
\frac{B_{4} }{2r} ), \eqnum{8}
\end{equation}
\begin{equation}
 p=\frac{\eta \,\cos \theta }{a} (B_{3} r+\frac{B_{4} }{r^{2} } ). \eqnum{9}
\end{equation}	 	
                              	                          
A substitution of these general equations in the edge conditions of system 
( 3 - 5 ) ( taking into account other conditions characterizing the 
electroosmosis process ) enables to obtain values of the fluid rate $I_v$ ,
electrochemical ion potentials $mu^\pm$ , the average electric field $E$ , the
electroosmotic transfer $P_e$ and the conductivity $K_\Sigma$ ( $I_E$ is
the electric current , $F$  is the Faraday constant ):
\begin{equation}
I_{V} =\left. \frac{\nu_{r} }{\cos\theta } \right| _{\,r=c}, \eqnum{10}
\end{equation}
\begin{equation}
E =\left.\frac{RT}{Fc\,\cos \theta }(\mu^{+} -\mu^{-} )\,\right | _{\,r=c}. \eqnum{11}
\end{equation}
\begin{equation}
P_{e} =I_{V} /I_{E},\eqnum{12}
\end{equation} 
\begin{equation}	 	
K_{\Sigma} =I_{E} /E.\eqnum{13}
\end{equation}		      	                                                   (13)
The edge problem is reduced to 12 linear algebraic equations by a 
substitution of the general solutions  ( 6 - 9 ) into the edge conditions 
of the system ( 3 - 5 ) ( taking into account other conditions characterizing 
the electroosmosis process ).
In the present paper the determination of values of the liquid velocity 
$\nu$ and electrochemical potentials  $mu^\pm$   was provided by solving this
system of  equations . The Gauss numerical method ,which is classified as 
direct , was used for the solution , taking into consideration relatively low 
order of the determinant [ 8 ]. The brief algorithm of the system solution is 
described below :
	                                                                  
1.Input  N - the order                         
of the equations system   
	
2.Formation of the              
augmented matrix                 
A ( N , N+1 )

3.Gauss method

4.Output of the solution X ( N )

Parameters for the problem solution were taken from the work [ 9 ] in which 
dispersion of synthetic diamond in HCl solution was investigated . Calculations
of $P_e$ and $K_\Sigma$  were provided for the  suspension with  particle
radius $a = 0.15\cdot  10^{-6}$ m , surface charge density 
$kju = 0.55\cdot 10^{-6}$ C/cm${^2}$ , electrolyte concentration $C_0 = 10^{-5}$ mole/l HCl .
The calculation results of the electroosmotic transfer $P_e$   and the conductivity
$K_\Sigma$   as functions of the volume particle fraction $\phi$ are given on
Fig.\ref{fig2}-\ref{fig3}.
$P_e$  was calculated for different non-dissolving boundary layer thicknesses
$h = 1$ nm  and $h = 10$ nm . The experimental dependencies of $P_e$  and
$K_\Sigma$  for the same values of particle radius $a$ , electrolyte concentration
$C_0$ , surface charge density  $Kju$ and volume particle fraction $\phi$  are
given on these figures too .

\section{Discussion}
	The dependence of the electroosmotic transfer $P_e$  on  the volume
particle fraction $\phi$ which was determined on the basis of the cell theory of 
the electroosmosis  shows the correspondence to the experimental regularity . 
The increase of  charged particle concentration in the suspension results in 
the increase of the degree of particle diffuse double layers' overlapping  and 
the augment of counterion  concentration in the porous space. Thus , the
electroosmotic transfer diminishes with the volume particle fraction increase .
Theoretically obtained function $P_e ( phi^{-1} )$ ( Fig.\ref{fig2} , line 1 for
$h = 1$ nm , line 2 for $h = 10$ nm ) is linear  which is typical for the
electroosmosis in concentrated disperse systems . The increase of the non-
dissolving boundary layer thickness leads to the decrease ( in these conditions ) 
of $P_e$ which is confirmed by the experiment ( Fig.\ref{fig2} , lines 3 , 4 ) .
The calculations of diamond particles dispersion conductivity $K_\Sigma$ versus
the volume particle fraction $\phi$ are shown on Fig.\ref{fig3} . Line 1 corresponds 
to experimental data , line 2 does to theoretical ones . In this case , when 
the electrolyte concentration is low and values of volume particle fraction are 
high , the effect of specific surface conductivity is significant . The 
resistance increase due to the increase of the insulating particle concentration 
is compensated and the conductivity of suspension is growing with the volume 
particle fraction augment ( so called capillary superconductivity [ 10 ] ).
	The calculation results ( accordingly the cell theory of the 
electroosmosis ) show the correspondence to and describe the main features of 
the experimental data : linear dependence of $P_e ( phi^{-1} )$ , the electroosmotic
transfer increase with the decrease of the non-dissolving boundary layer thickness , 
character of dependencies $K_\Sigma$   on $\phi$ .
The quantitative description of real disperse systems is impeded by a series 
of cases such as presence of particles of different sizes , deviation of 
particle shape from spherical , influence of specific surface conductivity . 
More detailed investigations of concentrated suspensions' properties  using 
the model systems which have more definite physical and chemical parameters 
are the subjects of future work . 
\section*{Acknowledgments}
The authors are grateful to Zharkih N. and Shilov V. for useful conversations .

\begin{figure}[tbp]
\caption{Structural elements in the cell model of a disperse system.
$a$ -radius of a particle; $h$ -thickness of the non-dissolving layer;
$b$ -radius of the non-dissolving layer in the cell; $c$ -cell radius;
 -particle surface charge ; 
2 -external boundary of the non-dissolving layer;
3 -external boundary of a diffuse layer.}
\label{fig1}
\end{figure}
\newpage
\begin{figure}[tbp]
\caption{Dependencies ofthe electroosmotic transfer $P_e$ on the reciprocal
volume particle fraction $j$.
1 -theoretical values for the non-dissolving boundary layer thickness $h = 1$ nm;
2 -theoretical values for the non-dissolving boundary layer thickness $h = 10$ nm;
3 -experimental data for system "diamond - HCL " in the presence of the 
saccharose solution which destructs the non-dissolving boundary layer;
4 -experimental data for system "diamond - HCL " without the saccharose
solution.}
\label{fig2}
\end{figure}
\newpage
\begin{figure}[tbp]
\caption{Dependence of the electrical conductivity $K_\Sigma$ on
the volume particle fraction $j$ .
1 -experimental data for system "diamond - HCL ";
2 -theoretical values.}
\label{fig3}
\end{figure}
\end{document}